\documentclass[prd,twocolumn,showpacs,floatfix,amsmath,nofootinbib,amssymb,floatfix]{revtex4}
\usepackage{graphicx,color,dcolumn,booktabs,bm}
\usepackage{longtable,lscape}
\usepackage{pdfpages}
\usepackage{txfonts}
\usepackage{overpic}
\usepackage{amssymb}
\usepackage{indentfirst}
\usepackage{feynmf}   %{feynmp}
\usepackage{slashed}  %for Feynman symbols
\usepackage{cases}
\usepackage{color}
\usepackage{multirow}
\usepackage{threeparttable}
\usepackage{epstopdf}
\usepackage{enumerate}
\usepackage{graphicx,color,dcolumn,booktabs,bm}
\usepackage[colorlinks, citecolor=blue,anchorcolor=red,menucolor=red, linkcolor=red,filecolor=red,urlcolor=blue,frenchlinks=red]{hyperref}

\graphicspath{{Figures/}} %

\begin{document}
%\begin{CJK}{GBK}{}

\title{New $\Omega_c^0$ baryons discovered by LHCb as the members of $1P$ and $2S$ states}
\author{Bing Chen$^{1,3}$}\email{chenbing@shu.edu.cn}
\author{Xiang Liu$^{2,3}$\footnote{Corresponding author}}\email{xiangliu@lzu.edu.cn}
\affiliation{$^1$Department of Physics, Anyang Normal University,
Anyang 455000, China\\$^2$School of Physical Science and Technology,
Lanzhou University,
Lanzhou 730000, China\\
%urlcolor=blue
$^3$Research Center for Hadron and CSR Physics, Lanzhou University
$\&$ Institute of Modern Physics of CAS,
Lanzhou 730000, China}

\date{\today}

\begin{abstract}
Inspired by the newly observed $\Omega_c^0$ states at LHCb, we decode their properties by performing an analysis of mass spectrum and decay behavior. Our studies show that the five narrow states, i.e., $\Omega_c(3000)^0$, $\Omega_c(3050)^0$, $\Omega_c(3066)^0$, $\Omega_c(3090)^0$, and $\Omega_c(3119)^0$, could be grouped into the $1P$ states with negative parity. Among them, the $\Omega_c(3000)^0$ and $\Omega_c(3090)^0$ states could be the $J^P=1/2^-$ candidates, while
$\Omega_c(3050)^0$ and $\Omega_c(3119)^0$ are suggested as the $J^P=3/2^-$ states. $\Omega_c(3066)^0$ could be regarded as a $J^P=5/2^-$ state. Since the the spin-parity, the electromagnetic transitions, and the possible hadronic decay channels $\Omega_c^{(\ast)}\pi$ have not been measured yet, other explanations are also probable for these narrow $\Omega_c^0$ states. Additionally, we discuss the possibility of the broad structure $\Omega_c(3188)^0$ as a $2S$ state with $J^P=1/2^+$ or $J^P=3/2^+$. In our scheme, $\Omega_c(3119)^0$ cannot be a $2S$ candidate.
\end{abstract}
 \maketitle

\section{Introduction}\label{sec1}

Establishing the higher radial and orbital excited states of heavy baryons is an interesting and important research issue of hadron spectroscopy in the past years~\cite{Chen:2016spr},
%(see the review paper \cite{Chen:2016spr} for the progress),
by which we can gain more information about non-perturbative behavior of quantum chromodynamics (QCD).
With the efforts of experiments, a significant progress has been made in searching for highly excited charmed baryons in the last few years. For examples, the $\Xi_c(2790)$, $\Xi_c(2815)$, and $\Xi_c(2980)$ have been measured by Belle with greater precision~\cite{Yelton:2016fqw}. A new decay mode, $D^+\Lambda$, was found for the $\Xi_c(3055)^+$ and $\Xi_c(3080)^+$~\cite{Kato:2016hca}. Recently, LHCb remeasured the resonance parameters of the $\Lambda_c(2880)^+$ and $\Lambda_c(2940)^+$~\cite{Aaij:2017vbw}, and confirmed the previously measured masses, decay widths, and the spin of $\Lambda_c(2880)^+$. In addition, a constraint on the spin-parity of the $\Lambda_c(2940)^+$ was given for the first time. More importantly, LHCb also observed a new broad state, $\Lambda_c(2860)^+$, with $J^P=3/2^+$~\cite{Aaij:2017vbw}.
Until now, these new measurements have greatly enriched the information of charmed baryons with the $cqq$ and $csq$ ($q~=$ $u$ or $d$ quark) configurations. It is obvious that the story of charmed baryons is still ongoing.

Very recently, the LHCb Collaboration again brought us a surprise due to the observation of five new narrow $\Omega_c^0$ states plus a broad structure in the $\Xi^+_cK^-$ invariant mass spectrum~\cite{Aaij:2017nav}.
%six $\Omega_c^0$ states in the $\Xi^+_cK^-$ invariant mass spectrum,
Here, the five narrow resonances are the $\Omega_c(3000)^0$, $\Omega_c(3050)^0$, $\Omega_c(3066)^0$, $\Omega_c(3090)^0$, and $\Omega_c(3119)^0$ states. Since this broad structure around 3188 MeV may be from a single resonance, or a superposition of several states, or other mechanism~\cite{Aaij:2017nav}, in this work we tentatively name this broad structure as the $\Omega_c(3188)^0$.
%and a broad structure around 3188 MeV (tentatively named as the $\Omega_c(3188)^0$ in this work).
The concrete experimental results of LHCb plus two established $\Omega_c^0$ baryons, $\Omega_c(2700)$ and $\Omega_c(2770)$~\cite{Olive:2016xmw}, are collected in Table \ref{table1} for the reader's convenience.
%We notice that the $\Omega_c(3000)^0$, $\Omega_c(3050)^0$, $\Omega_c(3066)^0$, $\Omega_c(3090)^0$, and $\Omega_c(3119)^0$ have the remarkable narrow widths.
These newly detected $\Omega_c^0$ states at LHCb not only make the $\Omega_c^0$ charmed baryon family become abundant, but also let us face how to categorize them into the $\Omega_c^0$ family.

\begin{table}[htbp]
\caption{The experimental results, including masses, widths, and decay modes, for the observed $\Omega_c^0$ states.} \label{table1}
\renewcommand\arraystretch{1.2}
\begin{tabular*}{85mm}{@{\extracolsep{\fill}}cccc}
\toprule[1pt]\toprule[1pt]
State~\cite{Aaij:2017nav,Olive:2016xmw} & Decay mode    & Mass (MeV)  & Width (MeV) \\
\midrule[0.8pt]
 $\Omega_c(2700)^0$ &             week                             & 2695.2$\pm$1.7                    &        \\
 $\Omega_c(2770)^0$ & $\Omega_c^0\gamma$        & 2765.9$\pm$2.0                    &         \\
 $\Omega_c(3000)^0$ & $\Xi_cK$                                & 3000.4$\pm$0.2$\pm$0.1$^{+0.3}_{-0.5}$    & 4.5$\pm$0.6$\pm$0.3         \\
 $\Omega_c(3050)^0$ & $\Xi_cK$                                & 3050.2$\pm$0.1$\pm$0.1$^{+0.3}_{-0.5}$    & 0.8$\pm$0.2$\pm$0.1        \\
 $\Omega_c(3066)^0$ & $\Xi_cK$, $\Xi^\prime_cK$   & 3065.6$\pm$0.1$\pm$0.3$^{+0.3}_{-0.5}$    & 3.5$\pm$0.4$\pm$0.2          \\
 $\Omega_c(3090)^0$ & $\Xi_cK$, $\Xi^\prime_cK$   & 3090.2$\pm$0.3$\pm$0.5$^{+0.3}_{-0.5}$    & 8.7$\pm$1.0$\pm$0.8           \\
 $\Omega_c(3119)^0$ & $\Xi_cK$, $\Xi^\prime_cK$   & 3119.1$\pm$0.3$\pm$0.9$^{+0.3}_{-0.5}$    & 1.1$\pm$0.8$\pm$0.4           \\
 $\Omega_c(3188)^0$ & $\Xi_cK$                                & 3188$\pm$5$\pm$13    & 60$\pm$15$\pm$11           \\
\bottomrule[1pt]\bottomrule[1pt]
\end{tabular*}
\end{table}

These new $\Omega_c^0$ excited states have aroused the theorists' great interests~\cite{Agaev:2017jyt,Karliner:2017kfm,Chen:2017sci,Wang:2017hej,Wang:2017vnc,Padmanath:2017lng,Cheng:2017ove,Wang:2017zjw,Zhao:2017fov,Yang:2017rpg,Huang:2017dwn,An:2017lwg,Kim:2017jpx,Kim:2017khv}. In the following, we briefly review the research status of them.
With the method of QCD sum rule, the masses of 2$S$ $\Omega_c^0$ states were calculated~\cite{Agaev:2017jyt}, where the assignment of $\Omega_c(3066)^0$ and $\Omega_c(3119)^0$ as the 2$S$ states were proposed. The strong decays of $P$-wave charmed baryons have been systematically investigated using the light-cone QCD sum rules~\cite{Chen:2017sci}, where the decay behaviors of the $\Omega_c(3000)^0$, $\Omega_c(3050)^0$, $\Omega_c(3066)^0$, $\Omega_c(3090)^0$, $\Omega_c(3119)^0$ were discussed. Based on the analysis from the Regge trajectories and the mass calculation via the heavy quark-light diquark model, the $\Omega_c(3090)^0$, and $\Omega_c(3119)^0$ were suggested as the 2$S$ states with $J^P=1/2^+$ and $J^P=3/2^+$, respectively~\cite{Cheng:2017ove}.
Within a constituent quark model, the analysis of strong behavior also favors the 2$S$ state assignment to the $\Omega_c(3119)^0$~\cite{Wang:2017hej}. This assignment to the $\Omega_c(3119)^0$ mentioned above is partly supported by the results in Refs.~\cite{Ebert:2011kk,Shah:2016nxi}, where the masses of 2$S$ $\Omega_c^0$ states were predicted to be around 3100 MeV. Different explanations have been proposed in Refs.~\cite{Karliner:2017kfm,Wang:2017vnc,Padmanath:2017lng,Wang:2017zjw}, where the $\Omega_c(3000)^0$, $\Omega_c(3050)^0$, $\Omega_c(3066)^0$, $\Omega_c(3090)^0$, $\Omega_c(3119)^0$ were suggested as the 1$P$  excited states with negative parity. In addition, some authors suggested that several newly observed narrow $\Omega_c^0$ states may be the charmed exotic systems~\cite{Yang:2017rpg,Huang:2017dwn,An:2017lwg,Kim:2017jpx,Kim:2017khv}.
%Obviously, these observed $\Omega_c^0$ states have inspired many theoretical interpretations.
Obviously, different conclusions were made by different theoretical methods (see Table II in Ref.~\cite{Wang:2017hej} for concise review). When facing such mess situation, more efforts are needed to these newly observed $\Omega_c^0$ baryons.

In this work, we give a systematic analysis for these new $\Omega_c^0$ states by performing the analysis of the mass spectra and especially the calculation of strong decays.
Firstly, we adopt a simple quark potential model to calculate the mass spectrum of low excited $\Omega_c^0$ states, where the heavy quark-light diquark picture is employed. Our study indicates that
$P$-wave $\Omega_c$ baryons have masses around 3.05 GeV, which overlaps with experimental observation. Thus, identifying five observed $\Omega_c$ state as \emph{P}-wave charmed baryons becomes possible. For further determining their spin-parity quantum numbers, we perform the study of their two-body Okubo-Zweig-Iizuka (OZI)-allowed strong decays under the quark pair creation (QPC) model, which may provide more abundant information of their inner structure.

The paper is organized as follows. After introduction, we give an analysis of mass spectrum of low-lying $\Omega_c^0$ excitations in the next section.
%By comparing theoretical and experimental results, the assignments to five narrow $\Omega_c^0$ states are determined.
In Sec. \ref{sec3}, we present an introduction of the QPC model for calculating the $\Omega_c^0$ baryon decays. In Sec. \ref{sec4}, we calculate the strong decays of the newly observed $\Omega_c^0$ baryon states and discuss our results. The paper ends with the summary and outlook in Sec. \ref{sec4}.

\section{An estimate of Masses of low-lying $\Omega_c^0$ states}\label{sec2}

%As a typical hadron composed of heavy and light quarks, the $\Omega_c^0$ baryon is an ideal system with the requirement of the heavy quark symmetry. In general, two strange quarks within the $\Omega_c^0$ system should decouple from the \emph{c} quark in the heavy quark limit. Under this scenario, the dynamics of a $\Omega_c^0$ state could be separated into two parts, $i.e.$, the degree of freedom between two $s$ quarks is usually called ``$\rho$ mode'', while the degree between the center of mass of two $s$ quarks and the $c$ quark is called ``$\lambda$ mode''. Usually, a three-body system can be treated as the interaction between light quark cluster and charm quark. If adopting the ordinary confining potential like the linear or harmonic form to depict this interaction, the mass of the $\rho$ mode excitations is higher than that of the $\lambda$ mode excitations~\cite{Copley:1979wj,Yoshida:2015tia}. Thus, experiment may first detect the $\lambda$ mode excited $\Omega_c^0$ states. Just considering this point, in this work, we mainly focus on the $\lambda$ mode excitations for these new $\Omega_c^0$ states.

Under the scenario of heavy quark symmetry\footnote{An $\Omega_c^0$ baryon resonance is composed of one heavy $c$ quark and two light $s$ quark, which could be regarded as a typical heavy-light hadron system. The heavy quark symmetry should be considered when one studys on its dynamics. In the heavy quark limit ($m_Q\rightarrow+\infty$), two strange quarks within the $\Omega_c^0$ system may decouple with the \emph{c} quark. In reality, the heavy quark symmetry is broken due to the finite charmed quark mass. However, we may expect that the heavy quark symmetry is a good approximation for the actual charmed hadron system. We need to mention that in this work we do not consider the $1/m_c$ correction.}, the dynamics of a $\Omega_c^0$ state could be separated into two parts, $i.e.$, the degree of freedom between two $s$ quarks is usually called ``$\rho$ mode'', while the degree of freedom between the center of mass of two $s$ quarks and the $c$ quark corresponds to ``$\lambda$ mode''. If adopting the ordinary confining potential like the linear or oscillator harmonic form to depict the interaction, the mass of the $\rho$ mode excitations is higher than that of the $\lambda$ mode excitations~\cite{Copley:1979wj,Yoshida:2015tia}. So experiments may first detect the $\lambda$ mode excited $\Omega_c^0$ states. Just considering this point, we firstly discuss whether these newly observed $\Omega_c^0$ states can be as the $\lambda$ mode excitations\footnote{In principle, a heavy-light baryon state should contain both $\rho$ mode and $\lambda$ mode. However, the investigation by Yoshida $et~al$. indicated that the $\rho$ and $\lambda$ modes are well separated for the charmed and bottomed baryons~\cite{Yoshida:2015tia}.}.
%Usually, a three-body system can be treated as the interaction between light quark cluster and charm quark.

Since the excitation between two $s$ quarks is not considered here, we may treat two $s$ quarks in these $\lambda$ mode excited $\Omega_c^0$ baryons as a special part with the antitriplet color structure and peculiar size, which is called as a light quark cluster or a light diquark. Under this heavy quark-light diquark picture, Isgur discussed the similarity of dynamics between heavy baryon and heavy-light meson~\cite{Isgur:1999rf}. Later, the heavy baryon masses were systematically calculated in such diquark picture~\cite{Ebert:2011kk,Chen:2014nyo}. In our recent work~\cite{Chen:2016iyi}, the properties of low-excited charmed and charmed-strange baryons have been described well under this diquark picture. We need to emphasize that the heavy quark-light diquark picture for the charmed baryon system is partially supported by the recent Belle's measurement of the production cross sections of charmed baryons~\cite{Niiyama:2017wpp}, where a factor of three difference for $\Lambda_c^+$ states over $\Sigma_c$ states was observed. This observation suggested a diquark structure in the ground state and low-lying excited $\Lambda_c^+$ baryons~\cite{Niiyama:2017wpp}.

For studying the dynamics of these low-lying $\Omega_c^0$ baryons, as described above, the light quark cluster (denoted as $\{ss\}$) could be an effective degree of freedom. Thus, a $\lambda$ mode excited $\Omega_c^0$ state may be simplified as a quasi-two-body system, which is similar to a meson system composed of a quark and an antiquark.
Here, the Cornell potential~\cite{Eichten:1978tg} is applied to phenomenologically describe confining part of interaction between a charm quark and an $\{ss\}$ diquark, $i.e.$,
%Since the $\Omega_c^0$ baryon has been simplified as a quasi-two-body system, we may use the following Cornell potential,
\begin{equation}
H^{conf}=-\frac{4}{3}\frac{\alpha_s}{r}+br+C. \label{eq1}
\end{equation}
%to describe the interaction between the light $\{ss\}$ cluster and charm quark.
The parameters, $\alpha_s$, $b$, and $C$ stand for the strength of color Coulomb
potential, the strength of linear confinement and a mass renormalized constant, respectively.  Furthermore, we expect that the following spin-dependent interaction terms which have been applied in studying the mass spectrum of different flavor mesons~\cite{Godfrey:1985xj} could also be suitable for the $\lambda$ mode excited $\Omega_c^0$ baryons. The first one is the magnetic-dipole-magnetic-dipole color hyperfine interaction
\begin{equation}
H_{hyp}=\frac{4}{3}\frac{\alpha_s}{m_lm_c}\left(\frac{8\pi}{3}\delta^3(\vec{r})\vec{s}_l\cdot\vec{s}_c+\frac{1}{r^3}\hat{S}_{lc}\right), \label{eq2}
\end{equation}
where the tensor operator is defined as
%$\hat{T}_{lc}=3\vec{s}_l\cdot\hat{r}\vec{s}_c\cdot\hat{r}-\vec{s}_l\cdot\vec{s}_c$.
\begin{equation}
\hat{S}_{lc}=\frac{3\vec{s}_l\cdot\vec{r}\vec{s}_c\cdot\vec{r}}{r^2}-\vec{s}_l\cdot\vec{s}_c.\nonumber
\end{equation}
Here $\vec{s}_l$ and $\vec{s}_c$ represent the spins of $\{ss\}$ diquark and $c$ quark, respectively. The second one corresponds to the spin-orbit interactions which contains two contributions. First piece is the color magnetic interaction due to one-gluon exchange,
\begin{eqnarray}
H^{cm}_{so}=\frac{4\alpha_s}{3r^3}\left(\frac{1}{m_l}+\frac{1}{m_c}\right)\left(\frac{\vec{s}_l}{m_l}+\frac{\vec{s}_c}{m_c}\right)\cdot\vec{L}. \label{eq3}
\end{eqnarray}
The second piece is the Thomas-precession term,
\begin{eqnarray}
H^{tp}_{so}=-\frac{1}{2r}\frac{\partial H^{conf}}{\partial r}\left(\frac{\vec{s}_l}{m^2_l}+\frac{\vec{s}_c}{m^2_c}\right)\cdot\vec{L}. \label{eq4}
\end{eqnarray}
With the expression of $H^{conf}$ in Eq.~(\ref{eq1}), the total spin-orbit interaction is given as
\begin{eqnarray}
H_{so}&=&\left[\left(\frac{2\alpha}{3r^3}-\frac{b}{2r}\right)\frac{1}{m_l^2}+\frac{4\alpha}{3r^3}\frac{1}{m_lm_c}\right]\vec{s}_l\cdot\vec{L}\nonumber\\&&
+\left[\left(\frac{2\alpha}{3r^3}-\frac{b}{2r}\right)\frac{1}{m_c^2}+\frac{4\alpha}{3r^3}\frac{1}{m_lm_c}\right]\vec{s}_c\cdot\vec{L}, \label{eq5}
\end{eqnarray}
Due to constraint of the Pauli exclusion principle, the total wave function of diquark $\{ss\}$ should be antisymmetric in the exchange of two strange quarks.
Since the spatial and flavor parts of this light quark cluster are always symmetric and color part is antisymmetric, the spin wave function should be symmetric, $i.e.$, $s_l=1$ for $\Omega_c^0$ state. In addition, $L$ denotes the orbital quantum number between light $\{ss\}$ cluster and charm quark. By solving Schr\"odinger equation, the masses of low-lying $\Omega_c^0$ states can be obtained directly. In our calculations, all spin-dependent interactions are treated as the leading-order perturbations for the orbital $\Omega_c^0$ excitations.

In our scheme, the mass matrix is calculated in the $jj$ coupling scheme. To this end, we adopt the basis $|s_l,L,j_l,s_c,J\rangle$, where $\vec{s}_l+\vec{L}=\vec{j}_l$ and $\vec{j}_l+\vec{s}_c=\vec{J}$. Five eigenvectors for the 1$P$ $\lambda$ mode excited $\Omega_c^0$ states include $|1,1,0,1/2,1/2\rangle\equiv|0,1/2^-\rangle$, $|1,1,1,1/2,1/2\rangle\equiv|1,1/2^-\rangle$, $|1,1,1,1/2,3/2\rangle\equiv|1,3/2^-\rangle$, $|1,1,2,1/2,3/2\rangle\equiv|2,3/2^-\rangle$, and $|1,1,2,1/2,5/2\rangle\equiv|2,5/2^-\rangle$, where the notation $|s_l,L,j_l,s_Q,J\rangle$ is abbreviated as $|j_l,J^P\rangle$ and the superscript $P$ denotes parity.

Due to the spin-orbit interaction $\vec{s}_c\cdot\vec{L}$ and the tensor interaction, two physical 1$P$ states with $J^P=1/2^-$ should be the mixtures of $|0,1/2^-\rangle$ and $|1,1/2^-\rangle$, which satisfy
\begin{eqnarray}
\begin{aligned}
 \left(
           \begin{array}{c}
                    |1P,1/2^-\rangle_L\\
                    |1P,1/2^-\rangle_H\\
                    \end{array}
     \right)&=\left(
           \begin{array}{cc}
                    \cos \theta_1  & -\sin\theta_1 \\
                    \sin\theta_1  & ~~\cos\theta_1\\
                    \end{array}
     \right)  \left(
           \begin{array}{c}
                     |0, 1/2^-\rangle \\
                     |1, 1/2^-\rangle \\
                    \end{array}
     \right). \label{eq6}
\end{aligned}
\end{eqnarray}
For the case of $J^P=3/2^-$ states, there exists
\begin{eqnarray}
\begin{aligned}
 \left(
           \begin{array}{c}
                     |1P,3/2^-\rangle_H\\
                     |1P,3/2^-\rangle_L\\
                    \end{array}
     \right)&=\left(
           \begin{array}{cc}
                    \cos\theta_2  & -\sin\theta_2 \\
                    \sin\theta_2  & ~~\cos\theta_2\\
                    \end{array}
     \right)  \left(
           \begin{array}{c}
                     |1, 3/2^-\rangle \\
                     |2, 3/2^-\rangle \\
                    \end{array}
     \right). \label{eq7}
\end{aligned}
\end{eqnarray}
The physical states with the same $J^P$ are distinguished by their different masses and widths. Here, the states with the lower and higher masses are denoted by the subscripts ``$L$'' and ``$H$'', respectively. Since the contribution from the contact hyperfine interaction (the first term of Eq.~(\ref{eq2})) is small for orbitally excited states, the spin-dependent interactions for the $P$-wave $\Omega_c^0$ baryons can be further simplified as
\begin{equation}
H_S=V_l\vec{s}_l\cdot\vec{L}+V_c\vec{s}_c\cdot\vec{L}+V_t\hat{S}_{lc}. \label{eq8}
\end{equation}
By the comparison between Eq.~(\ref{eq2}) and Eq.~(\ref{eq5}), we may define the expressions of $V_l$, $V_c$, and $V_t$, $i.e.$,
\begin{equation}
\begin{split}
&V_l = \left(\frac{2\alpha}{3r^3}-\frac{b}{2r}\right)\frac{1}{m_l^2}+\frac{4\alpha}{3r^3}\frac{1}{m_lm_c},~~~~\\&
V_c = \left(\frac{2\alpha}{3r^3}-\frac{b}{2r}\right)\frac{1}{m_c^2}+\frac{4\alpha}{3r^3}\frac{1}{m_lm_c},~~~~\\&
V_t = \frac{4}{3}\frac{\alpha_s}{m_lm_c}\frac{1}{r^3}.\nonumber
\end{split}
\end{equation}
For \emph{P}-wave states with $J^P=1/2^-$, the mass matrix is
\begin{eqnarray}
\langle\Phi_{1/2}\mid H_S\mid\Phi_{1/2}\rangle=\left(
           \begin{array}{ccc}
                    -2V_l-\frac{4}{3}V_t  & -\frac{3V_c+V_t}{3\sqrt{2}}\\
                    -\frac{3V_c+V_t}{3\sqrt{2}}  & -V_l-\frac{1}{2}V_c+\frac{1}{3}V_t \\
                    \end{array}
     \right).\nonumber
\end{eqnarray}
Similarly, for two states with $J^P=3/2^-$, we have
\begin{eqnarray}
\begin{split}
\langle\Phi_{3/2}\mid H_S\mid\Phi_{3/2}\rangle=\left(
           \begin{array}{ccc}
                    \frac{1}{4}V_c-V_l+\frac{5}{6}V_t  & \frac{4V_t-15V_c}{12\sqrt{5}}\\
                    \frac{4V_t-15V_c}{12\sqrt{5}}  & V_l-\frac{3}{4}V_c-\frac{1}{30}V_t \\
                    \end{array}
     \right).\nonumber
\end{split}
\end{eqnarray}
For the $J^P=5/2^-$ state, we get
\begin{equation}
\begin{split}
\langle 2, 5/2^- \mid H_S\mid 2, 5/2^-\rangle =V_l+\frac{1}{2}V_c-\frac{1}{5}V_{t}.\nonumber
\end{split}
\end{equation}
The notations $\mid\Phi_{1/2}\rangle$ and $\mid\Phi_{3/2}\rangle$ in the above expressions have the definition
\begin{eqnarray}
\begin{split}
\mid\Phi_J\rangle
=\left(
           \begin{array}{cc}
                    \mid j_l=J-1/2, J^P\rangle  \\
                    \mid j_l=J+1/2, J^P\rangle  \\
                    \end{array}
     \right). \nonumber
\end{split}
\end{eqnarray}

There are five parameters, $m_c$, $m_l$, \emph{b}, $\alpha_s$, $\sigma$, and $C$, in this nonrelativistic quark potential model. The $c$ quark mass is taken to be $m_c=1.68$ GeV from our previous work~\cite{Chen:2016iyi}. Based on the SU(3) flavor symmetry, there exists the similarity of the dynamics for $\Sigma_c$ and $\Omega_c$ baryon families. The averaged mass of the ground $\Sigma_c$ states ($\Sigma_c(2455)$ and $\Sigma_c(2520)$) is about 2496 MeV, while the averaged mass of the ground $\Omega_c$ states ($\Omega_c(2700)$ and $\Omega_c(2770)$) is about 2743 MeV. Thus, the splitting of the ground $\Sigma_c$ and $\Omega_c$ states is given by
\begin{equation}
\bar{M}_{[\Sigma_C(2455),~\Sigma_C(2520)]}-\bar{M}_{[\Omega_C(2700),~\Omega_C(2770)]}\approx 247~\rm {MeV}, \label{eq9}
\end{equation}
which could be regarded as the mass difference between the light quark clusters of $\{uu\}$ and $\{ss\}$~\cite{Jaffe:2004ph}. Then, we evaluate the mass of the axial-vector light quark cluster $m_l$ as 0.91 GeV since the mass of $\{uu\}$ cluster in $\Sigma_c$ baryons has been fixed as 660 MeV by a relativistic flux tube model in our previous work~\cite{Chen:2014nyo}.
%, by the following relation
%\begin{equation}
%\frac{4\times m_{\Omega_c(2770)^0}+2\times m_{\Omega_c(2700)^0}}{6}-m_{\Lambda_c(2286)^+}\approx 456~\rm {MeV}. \label{eq10}
%\end{equation}
%Since the mass of scalar diquark $[qq]$ in $\Lambda_c^+$ baryons was fixed as 451 MeV by a relativistic flux tube model~\cite{Chen:2014nyo}, the mass of light diquark $\{ss\}$ in $\Omega_c^0$ baryons is evaluated about 0.91 GeV.
The strength of color Coulomb potential $\alpha_s$ and the strength of linear confinement $b$ are taken as 0.34 and 0.120 GeV$^{-1}$, respectively, which can fairly reproduce the averaged masses of the observed 1$S$, 1$P$, and 2$S$ $\Omega_c^0$ candidates. To reproduce the mass splitting of the $\Omega_c(2700)^0$ and $\Omega_c(2770)^0$, the parameter $\sigma$ is fixed as 1.00. Finally, the constant $C$ is determined as 0.16 GeV.

\begin{figure}[htbp]
\begin{center}
\includegraphics[width=8.6cm,keepaspectratio]{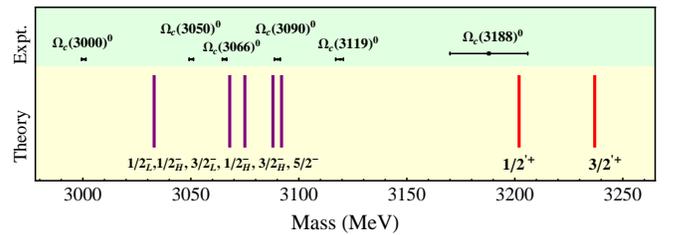}
\caption{A comparison of the predicted 1$P$ and 2$S$ $\Omega_c^0$  masses with the
newly observed states by LHCb.}\label{Fig1}
\end{center}
\end{figure}

With the parameters above, the masses of two ground (1$S$) $\Omega_c$ states are predicted as 2698 MeV and 2765 MeV which are in good agreement with the experimental results (see Table~\ref{table1}). For a clear  comparison, we collect the predicted 1$P$ and 2$S$ $\Omega_c^0$  masses and the newly observed states by LHCb together in Fig.~\ref{Fig1}. According to the predicted masses, we find:
\begin{itemize}

\item Among the five 1$P$ states, a $J^P=1/2^-$ state should have the lowest mass. This conclusion is supported by most theoretical works~\cite{Karliner:2017kfm,Wang:2017vnc,Padmanath:2017lng,Roberts:2007ni,Wang:2017zjw,Maltman:1980er,Yoshida:2015tia}. Then we may conjecture that $\Omega_c(3000)^0$ is a good candidate for the $|1P,1/2^-\rangle_L$ state if we try to assign these 5 narrow states to the 1$P$ $\Omega_c^0$ family.

\item The masses of the $|1P,1/2^-\rangle_H$ state, two $3/2^-$ states, and one $5/2^-$ state are predicted in the range $3068\sim3092$ MeV. Considering the intrinsic uncertainties of the quark potential models, we cannot determine the quantum numbers for the $\Omega_c(3050)^0$, $\Omega_c(3066)^0$, $\Omega_c(3090)^0$, and $\Omega_c(3119)^0$ states only by the analysis of the mass spectrum mentioned above.

\item The broad structure, the $\Omega_c(3188)^0$, may be regarded as an $2S$ state with $J^P=1/2^+$ or $3/2^+$, or
their overlapping structure.

\end{itemize}

With the mass matrices above, the mixing angles in Eqs.~(\ref{eq6}) and ~(\ref{eq7}) are obtained as $\theta_1=158^\circ$ and $\theta_2=159^\circ$, respectively. However, the spin-dependent interactions for the excited $\Omega_c^0$ states have never been understood well in three-body picture. We should treat these obtained mixing angles with some care.
Although the spin-parity quantum numbers of $\Omega_c(3050)^0$, $\Omega_c(3066)^0$, $\Omega_c(3090)^0$, and $\Omega_c(3119)^0$
cannot be fixed under the simplified quark potential model in the diquark picture, we may conclude that
the $1P$ states of $\Omega_c$ baryon family have mass around 3.05 GeV which overlaps with the
the masses of these observed states. Thus, assigning $\Omega_c(3050)^0$, $\Omega_c(3066)^0$, $\Omega_c(3090)^0$, and $\Omega_c(3119)^0$ states as $1P$ states of $\Omega_c$ family is still possible. Our tentative conclusion is also supported by the result given by Lattice QCD recently~\cite{Padmanath:2017lng}.

In the following section, we will employ the QPC model to give further constraints for the assignments of these narrow $\Omega_c^0$ states by analyzing their strong decays.

\section{The $\Omega_c^0$ baryon decays in the QPC model}\label{sec3}

The idea of QPC strong decay model was starting from Micu~\cite{Micu:1968mk}, Carlitz and Kislinger~\cite{Carlitz:1970xb}, and later formulated by the Orsay group~\cite{LeYaouanc:1972vsx,LeYaouanc:1988fx}. For an OZI-allowed decay process of a hadron system, the QPC model suggests that a quark-antiquark pair is created from the vacuum and then regroups into two outgoing hadrons by a quark rearrangement process. Thus the $q\bar{q}$ pair shall carry the quantum number of $0^{++}$, suggesting that they are in a $^3P_0$ state. So the QPC model is also named ``the $^3P_0$ model''. This model has been successfully applied to study of different kinds of hadron systems for their OZI-allowed strong decays. Here we just quite some works which have been devoted to the strong decay behaviors of excited heavy baryons. With the QPC model, the strong decays of the $S$-wave, $P$-wave, $D$-wave, and radially excited charmed baryons have been studied~\cite{Chen:2007xf}. The QPC model was taken to study the decay processes of ground and excited bottom baryons~\cite{Limphirat:2010zz}. Recently, the decays of $P$-wave excitations of charmed strange baryons have been studied systematically by the QPC model~\cite{Ye:2017yvl}.

\begin{figure}[htpb]
\begin{center}
\includegraphics[width=8.6cm,keepaspectratio]{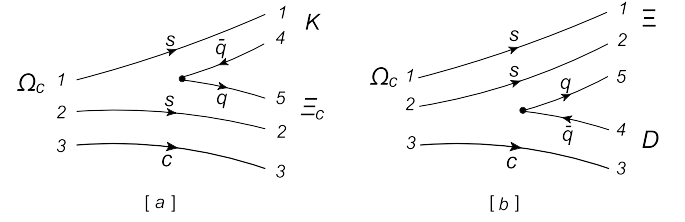}
\caption{A diagram for the decay processes, $\Omega_c^0\rightarrow\Xi_cK$ and $\Omega_c^0\rightarrow\Xi{D}$, where the quarks have been numbered.}\label{Fig2}
\end{center}
\end{figure}

According to the excited energy, there are two kinds of decay processes for an excited charmed baryon state (see Fig.~\ref{Fig2}). When the mass of a $\Omega_c^0$ excitation is higher than about 2.97 GeV, the threshold of $\Xi_cK$ channel, only the left process (labelled by $[a]$) is allowed. When the mass is higher the threshold of $D\Xi$ channel, then two kinds of decay processes depicted in Fig.~\ref{Fig2} are possible. We now take the left process, $\Omega_c(A)\rightarrow{\Xi_c(B)}+K(C)$, to show how to obtain the partial wave amplitudes. To describe the decay process, the transition operator $\mathcal {\hat{T}}$ of the $^3P_0$ model is given by
\begin{equation}
\begin{split}
\mathcal {\hat{T}}=&-3\gamma
\sum_{\text{\emph{m}}}\langle1,m;1,-m|0,0\rangle \iint
d^3\vec{k}_4d^3\vec{k}_5\delta^3(\vec{k}_4+\vec{k}_5)\\ &\times\mathcal
{Y}_1^m\left(\frac{\vec{k}_4-\vec{k}_5}{2}\right)\omega^{(4,5)}\varphi^{(4,5)}_0\chi^{(4,5)}_{1,-m}d^\dag_{4}(\vec{k}_4)d^\dag_{5}(\vec{k}_5),% \nonumber
\end{split}\label{eq10}
\end{equation}
in a non-relativistic limit. Here the $\omega_0^{(4,5)}$ and $\varphi^{(4,5)}_0$ are the color and flavor wave functions of the $\bar{q}_4q_5$ pair created from the vacuum. Therefore, $\omega^{(4,5)}=(R\bar{R}+G\bar{G}+B\bar{B})/\sqrt{3}$ and $\varphi^{(4,5)}_0=(u\bar{u}+d\bar{d}+s\bar{s})/\sqrt{3}$ are color and flavor singlets. The $\chi^{(4,5)}_{1,-m}$ represents the pair production in a spin
triplet state. The solid harmonic polynomial $\mathcal {Y}_1^m(\vec{k})\equiv|\vec{k}|\mathcal {Y}_1^m(\theta_k,\phi_k)$ reflects the momentum-space distribution of the $\bar{q}_4q_5$. The dimensionless parameter $\gamma$ describes the strength of the quark-antiquark pair created from the vacuum. The value of $\gamma$ is usually fixed as a constant by fitting the well measured partial decay widths.

When the mock state~\cite{Hayne:1981zy} is adopted to describe the spatial wave function of a hadron state, the helicity amplitude can be easily constructed in the $JJ$ basis. The mock state for the initial state $A$ is given by

\begin{equation}
\begin{split}
|A({n_A}^{2S_A+1}&L_A^{J_Aj_A}(\vec{P}_A)\rangle\equiv\\
&\omega_A^{123}\phi_A^{123}\prod_A\int d^3\vec{k}_1d^3\vec{k}_2d^3\vec{k}_3\delta^3(\vec{k}_1+\vec{k}_2+\vec{k}_3-\vec{P}_A)\\
&\times\Psi_{n_A}^{L_Al_A}(\vec{k}_1,\vec{k}_2,\vec{k}_3)|q_1(\vec{k}_1)q_2(\vec{k}_2)q_3(\vec{k}_3)\rangle,
\end{split}\label{eq11}
\end{equation}
where the $\omega_A^{123}$ and $\phi_A^{123}$ are the color and flavor wave functions of baryon $A$. The wave function of a $\Xi_c$ baryon in Fig.~\ref{Fig3} can be constructed in the same way. The wave function of the \emph{K} meson is
\begin{equation}
\begin{split}
|C({n_C}^{2S_C+1}&L_C^{J_Cj_C}(\vec{P}_C)\rangle\equiv\omega_C^{14}\phi_C^{14}\prod_C \int d^3\vec{k}_1d^3\vec{k}_4\\
&\times\delta^3(\vec{k}_1+\vec{k}_4-\vec{P}_C)~\psi_{n_C}^{L_Cl_C}(\vec{k}_1,\vec{k}_4)|q_1(\vec{k}_1)\bar{q}_4(\vec{k}_4)\rangle.
\end{split}\label{eq12}
\end{equation}
Here, the symbols of $\prod_i~(i=A, B$, and $C)$ represent the Clebsch-Gordan coefficients for the initial and final hadrons, which arise from the couplings among the orbital, spin, and total angular momentum and their projection of $l_z$ and $s_z$ to $j_z$. More specifically, $\prod_i~(i=A, B$, and $C)$ are
\begin{equation}
\begin{split}
&\langle s_1m_1, s_2m_2|s_{12}m_{12}\rangle\langle s_{12}m_{12}, L_Al_A|J^l_Aj^l_A\rangle\langle J^l_Aj^l_A, s_3m_3|J_Aj_A\rangle,\\&
\langle s_2m_2, s_5m_5|s_{25}m_{25}\rangle\langle s_{25}m_{25}, L_Bl_B|J^l_Bj^l_B\rangle\langle J^l_Bj^l_B, s_3m_3|J_Bj_B\rangle,\\&
\langle s_1m_1, s_4m_4|S_Cs_C\rangle\langle L_Cl_C, S_Cs_C|J_Cj_C\rangle,\nonumber
\end{split}
\end{equation}
respectively. $J^l_\alpha$ ($\alpha=A$ or $B)$ refers to the total angular momentum of the light diquark in the heavy baryons. To obtain the analytical amplitudes, the following simple harmonic oscillator (SHO) wave function is usually employed to construct the spatial wave function of the hadron state,
\begin{equation}
\begin{split}
&\psi^n_{Lm}(\textbf{p}) \\&=\frac{(-1)^n}{\beta^{3/2}}\sqrt{\frac{2(2n-1)!}{\Gamma(n+L+\frac{1}{2})}}\left(\frac{p}{\beta}\right)^L
e^{-\frac{p^2}{2\beta^2}}L^{L+1/2}_{n-1}\left(\frac{p^2}{\beta^2}\right)\mathcal
{Y}_{Lm}(\textbf{p}),
\end{split}\label{eq13}
\end{equation}
where $L^{L+1/2}_{n-1}\left(p^2/\beta^2\right)$ is an associated Laguerre polynomial. The helicity amplitude $\mathcal {M}^{j_A,j_B,j_C}(q)$ is defined by
\begin{eqnarray}\label{eq14}
\langle BC|\mathcal {\hat{T}}|A\rangle=
\delta^3(\vec{P}_A-\vec{P}_B-\vec{P}_C)\mathcal
{M}^{j_A,j_B,j_C}(q).
\end{eqnarray}
Here \emph{q} represents the momentum of an outgoing meson in the rest frame of a meson \emph{A}, which is given by
\begin{equation}\label{eq15}
q=\frac{\sqrt{\left[M_A^2-(M_B+M_C)^2\right]\left[M_A^2-(M_B-M_C)^2\right]}}{2M_A}.
\end{equation}
For comparison with experiments, one needs to obtain the partial wave amplitudes $\mathcal {M}_{LS}(q)$ via the Jacob-Wick formula~\cite{Jacob:1959at}
\begin{equation}
\begin{aligned}\label{eq16}
\mathcal {M}_{LS}(q)=&\frac{\sqrt{2L+1}}{2J_A+1}\\
&\times\sum_{\text{$j_B$,$j_C$}}\langle L0Jj_A|J_Aj_A\rangle\langle J_Bj_B,J_Cj_C|Jj_A\rangle\mathcal
{M}^{j_A,j_B,j_C}(q).
\end{aligned}
\end{equation}

Finally, the decay width $\Gamma(A\rightarrow BC)$ is derived analytically in terms of the partial wave amplitudes in the $A$ rest frame,
\begin{equation}
\begin{aligned}\label{eq17}
\Gamma(A\rightarrow BC)=2\pi\frac{E_BE_C}{M_A}q\sum_{L,S}|\mathcal
{M}_{LS}(q)|^2.
\end{aligned}
\end{equation}

With Eqs.~(\ref{eq10}), (\ref{eq11}), (\ref{eq12}), (\ref{eq14}), and (\ref{eq16}), the full expression of $\mathcal {M}_{LS}(q)$ in the rest frame of the baryon \emph{A} is
\begin{widetext}
\begin{equation}
\begin{aligned}
\mathcal {M}_{LS}(q)=&-3\gamma\xi\frac{\sqrt{2L+1}}{2J_A+1}\sum_{l_i,m_j}\langle L0;Jj|J_Aj_A\rangle\langle J_Bj_B;J_Cj_C|Jj\rangle\langle s_1m_1;s_2m_2|s_{dA}m_{12}\rangle\langle s_{dA}m_{12};L_Al_A|J^l_Aj^l_A\rangle\langle J^l_Aj^l_A;s_3m_3|J_Aj_A\rangle \\
&\times\langle s_2m_2;s_5m_5|s_{dB}m_{25}\rangle\langle s_{dB}m_{25};L_Bl_B|J^l_Bj^l_B\rangle\langle J^l_Bj^l_B;s_3m_3|J_Bj_B\rangle\langle s_1m_1;s_4m_4|S_Cs_C\rangle\langle S_Cs_C;L_Cl_C|J_Cj_C\rangle \\
&\times\langle s_4m_4;s_5m_5|1-m\rangle\langle1,m;1,-m|0,0\rangle\langle\omega^{235}_B\omega^{14}_C|\omega^{45}_0\omega^{123}_A\rangle\int\cdots\int d^3\vec{k}_1\cdots d^3\vec{k}_5\delta^3(\vec{k}_1+\vec{k}_2+\vec{k}_3)\\
&\times\delta^3(\vec{q}-\vec{k}_1-\vec{k}_4)\delta^3(\vec{k}_4+\vec{k}_5)\delta^3(\vec{q}+\vec{k}_2+\vec{k}_3+\vec{k}_5)\Psi_A(\vec{k}_1,\vec{k}_2,\vec{k}_3)\Psi^*_B(\vec{k}_1,\vec{k}_2,\vec{k}_4)\psi^*_C(\vec{k}_3,\vec{k}_5)\mathcal
{Y}_1^m\left(\frac{\vec{k}_4-\vec{k}_5}{2}\right),
\label{eq18}
\end{aligned}
\end{equation}
\end{widetext}
where, $i=A, B, C$ and $j=1, 2, \cdots, 5$. The color matrix element $\langle\omega^{235}_B\omega^{14}_C|\omega^{45}_0\omega^{123}_A\rangle$ is a constant which can be absorbed into the parameter $\gamma$. More details for deducing these flavor matrix elements, $\xi=\langle \varphi^{235}_B\varphi^{14}_C|\varphi^{45}_0\varphi^{123}_A\rangle$, will be presented in the Appendix~\ref{A}. In addition, we collect these obtained partial wave amplitudes in Appendix~\ref{B}.

\section{Results and discussions}\label{sec4}

%\subsection{Two $1/2^-$ states}\label{a}

As stressed here, an important motivation of this work is to test whether the five narrow $\Omega_c^0$ states can be grouped into the 1$P$ family or not. The primary challenge to this question is to explain their narrow widths together. According to our result given by the potential model, the $\Omega_c(3000)^0$ state could be regarded as a $|1P,1/2^-\rangle_L$ candidate with the predominant component of $|0, 1/2^-\rangle$. If this is true, however, $\Omega_c(3000)^0$ might be a broad state since it can decay into $\Xi_cK$ through $S$-wave channel with large phase space. The results given in Refs.~\cite{Wang:2017hej,Cheng:2017ove,Zhao:2017fov} confirmed this point. Thus $\Omega_c(3000)^0$ should has the predominant component of $|1, 1/2^-\rangle$. If this is true, the narrow width of $\Omega_c(3000)^0$ can be understood since the decay process of $\Xi_cK$ is forbidden for the component of $|0, 1/2^-\rangle$. Then the decay of $\Omega_c(3000)^0$ is strongly suppressed due to the mixing effect. According to Eq.~(\ref{eq6}), we have
\begin{eqnarray}
\begin{aligned}
 \left(
           \begin{array}{c}
                    |\Omega_c(3000)^0\rangle\\
                    |\Omega_c(X)^0\rangle\\
                    \end{array}
     \right)&=\left(
           \begin{array}{cc}
                    \cos \theta_1  & -\sin\theta_1 \\
                    \sin\theta_1  & ~~\cos\theta_1\\
                    \end{array}
     \right)  \left(
           \begin{array}{c}
                     |0, 1/2^-\rangle \\
                     |1, 1/2^-\rangle \\
                    \end{array}
     \right). \label{eq19}
\end{aligned}
\end{eqnarray}
Then a question arises that which state among $\Omega_c(3050)^0$, $\Omega_c(3066)^0$, $\Omega_c(3090)^0$, and $\Omega_c(3119)^0$, can be the mixture partner of $\Omega_(3000)^0$. To solve this problem, we calculate the ratio of their decay widths with respect to the width of $\Omega_c(3000)^0$. The advantage of this way is that one can avoid the uncertainty of the phenomenological parameter $\gamma$ in the QPC model. With the decay amplitudes listed in Appendix~\ref{B}, firstly, we fix the mixing angle $\theta_1$ directly. Furthermore, we obtain the value of $\gamma$ by reproducing their measured decay widths. The results of $\theta_1$ and $\gamma$ are collected in Table~\ref{table2}.

\begin{table}[htbp]
\caption{As the mixture partner of $\Omega_c(3000)^0$ (Eq.~\ref{eq19}), the values of $\gamma$ and $\theta_1$ obtained by the ratios of decay widths of $\Omega_c(3050)^0$, $\Omega_c(3066)^0$, $\Omega_c(3090)^0$, and $\Omega_c(3119)^0$ to $\Omega_c(3000)^0$.}\label{table2}
\renewcommand\arraystretch{1.2}
\begin{center}
\begin{tabular*}{85mm}{c@{\extracolsep{\fill}}cccc}
\toprule[1pt]\toprule[1pt]
 $\Omega_c(X)^0$    & $\Omega_c(3050)^0$     &  $\Omega_c(3066)^0$      &  $\Omega_c(3090)^0$       &  $\Omega_c(3119)^0$       \\
\toprule[1pt]
 $\gamma$           & $0.208$                & $0.336$                  & $0.469$                   & $-$            \\
 $\theta_1$         & $151.8^\circ$          & $128.2^\circ$            & $116.3^\circ$             & $-$      \\
\bottomrule[1pt]\bottomrule[1pt]
\end{tabular*}
\end{center}
\end{table}

If $\Omega_c(3050)^0$ is the partner of $\Omega_c(3000)^0$, the fixed mixing angle $\theta_1$ is about $151.8^\circ$ which is comparable with the obtained value by quark potential model. But the value of $\gamma$ is only about 0.208 which is too small to reproduce the widths of other three states. The case of $\Omega_c(3066)^0$ is alike. We find that the $\Omega_c(3119)^0$ can not be regarded as the partner of $\Omega_c(3000)^0$ in our scheme. Indeed only $\Omega_c(3090)^0$ could be the candidate as the partner of $\Omega_c(3000)^0$. The angle $\theta_1=116.3^\circ$ fixed by $\Omega_c(3000)^0$ and $\Omega_c(3090)^0$ indicates that the quark potential model overestimate the mixing of two $1/2^-$ $\Omega_c^0$ states. As shown later, the value of $\gamma=0.469$ can also naturally reproduce the widths of $\Omega_c(3050)^0$, $\Omega_c(3066)^0$, and $\Omega_c(3119)^0$.

\begin{table}[htbp]
\caption{The partial widths of $\Omega_c(3050)^0$, $\Omega_c(3066)^0$, and $\Omega_c(3119)^0$ with the $|1,3/2^-\rangle$, $|2,3/2^-\rangle$, and $|2,5/2^-\rangle$ assignment. The measured widths in MeV in square brackets are listed for comparison.}\label{table3}
\renewcommand\arraystretch{1.2}
\begin{center}
\begin{tabular*}{85mm}{c@{\extracolsep{\fill}}cccccc}
\toprule[1pt]\toprule[1pt]
      \multirow{2}{*}{Assignment}   &     \multicolumn{2}{c}{$\Omega_c(3050)^0$}     & \multicolumn{2}{c}{$\Omega_c(3066)^0$}  &  \multicolumn{2}{c}{$\Omega_c(3119)^0$}  \\
                                    & \multicolumn{2}{c}{$[0.8\pm0.3]$ } & \multicolumn{2}{c}{$[3.5\pm0.6]$ } & \multicolumn{2}{c}{$[1.1\pm1.2]$ } \\
\toprule[1pt]
 $|1,3/2^-\rangle$  & $\Xi_cK$                   & $\times$      & $\Xi_cK$                   & $\times$     & $\Xi_cK$                   & $\times$    \\
                                    &                            &                &                            &              & $\Xi^\prime_cK$            &  0.2   \\
 $|2,3/2^-\rangle$  & $\Xi_cK$                   & 1.8           & $\Xi_cK$                   & 2.4          & $\Xi_cK$                   &  4.8    \\
                                    &                            &                &                            &              & $\Xi^\prime_cK$            &  0.3   \\
 $|2,5/2^-\rangle$  & $\Xi_cK$                   & 1.8           & $\Xi_cK$                   & 2.4          & $\Xi_cK$                   &  4.8    \\
                                    &                            &                &                            &              & $\Xi^\prime_cK$            &  0.1   \\
\bottomrule[1pt]\bottomrule[1pt]
\end{tabular*}
\end{center}
\end{table}

In the following, we further study the $\Omega_c(3050)^0$, $\Omega_c(3066)^0$, $\Omega_c(3119)^0$, and try to find which one is the most likely $5/2^-$ state. To this end, we calculate their partial widths with the $|1,3/2^-\rangle$, $|2,3/2^-\rangle$, and $|2,5/2^-\rangle$ assignments. All results are listed in Table~\ref{table3} where the $\gamma=0.469$ is used. As a $|2,5/2^-\rangle$ state, we find that the predicted widths of $\Omega_c(3050)^0$ and $\Omega_c(3119)^0$ are about $2\sim4$ times larger than the center values of the experimental widths, which implies that the possibility of $\Omega_c(3050)^0$ and $\Omega_c(3119)^0$ as the $|2,5/2^-\rangle$ state could be preliminarily excluded. The predicted width of $\Omega_c(3066)^0$ is about 2.4 MeV which is compatible with experimental result (see Table~\ref{table3}). So $\Omega_c(3066)^0$ is the most likely of the $J^P=5/2^-$ candidate. We may stress that the mass of the $1P$ $ 5/2^-$ $\Omega_c^0$ state was also predicted around 3.05 GeV in Refs.~\cite{Maltman:1980er,Ebert:2011kk,Yoshida:2015tia}. So the assignment of $\Omega_c(3066)^0$ as a $5/2^-$ state is probable.

\begin{figure}[htbp]
\begin{center}
\includegraphics[width=8.6cm,keepaspectratio]{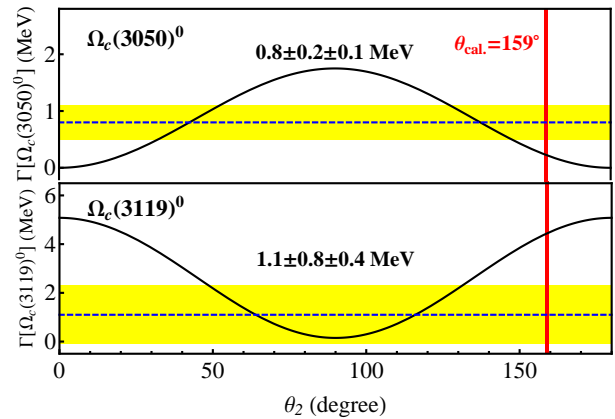}
\caption{The dependence of the total decay widths of the $\Omega_c(3050)^0$ and $\Omega_c(3119)^0$ states on the mixing angle $\theta_2$ in Eq.~(\ref{eq7}). Here, the $\Omega_c(3050)^0$ and $\Omega_c(3119)^0$ are treated as a $|1P,3/2^-\rangle_L$ and $|1P,3/2^-\rangle_H$ states, respectively. The experimental widths are also presented for comparison. }\label{Fig3}
\end{center}
\end{figure}

To investigate the possibility of $\Omega_c(3050)^0$ and $\Omega_c(3119)^0$ as two mixtures of  $|1, 3/2^-\rangle$ and $|2, 3/2^-\rangle$ (see Eq.~(\ref{eq2})), we study their decay behaviors in Fig. \ref{Fig3} where the dependence of their total decay widths on the mixing angle $\theta_2$ is illustrated. The mixing angle $\theta_2$ given by quark potential model seems also be overestimated since the predicted width of $\Omega_c(3119)^0$ is still larger than experiment. If we choose a suite mixing angle, $e.g.$, $\theta_2\approx130^\circ$, the calculated widths are about 1.0 MeV and 2.2 MeV for $\Omega_c(3050)^0$ and $\Omega_c(3119)^0$, respectively. Obviously, the measured decay widths of $\Omega_c(3050)^0$ and $\Omega_c(3119)^0$ could be understood in the mixing scenario. In a word, the $\Omega_c(3050)^0$ and $\Omega_c(3119)^0$ are suggested to be the $|1P, 3/2^-\rangle_L$ and $|1P, 3/2^-\rangle_H$ candidates in our scheme.

Besides these five narrow $\Omega_c^0$ states, LHCb also reported a broad structure around 3188 MeV, denoted as the $\Omega_c(3188)^0$, in the $\Xi_c^+K^-$ invariant mass distribution~\cite{Aaij:2017nav}. As shown in Fig. \ref{Fig1}, the $\Omega_c(3188)^0$ can be grouped into the 2$S$ family with $J^P=1/2^+$ or $3/2^+$. The comparable predicted masses for the 2$S$ $\Omega_c^0$ states were also obtained in Refs.~\cite{Shah:2016mig,Roberts:2007ni,SilvestreBrac:1996bg,Valcarce:2008dr}. To further check this possibility, we calculated the partial and total decay widths of two 2\emph{S} $\Omega_c^0$ states where the predicted masses were used. As shown in Table \ref{table4}, the largest decay channel for the $2S$ $\Omega_c^0$ states is $\Xi_cK$, which may explain why the $\Omega_c(3188)^0$ was first observed in this channel. Nevertheless, we notice that theoretical total decay widths of 2$S$ $\Omega_c^0$ states with $J^P=1/2^+$ and $J^P=3/2^+$ are only 13.8 MeV and 13.1 MeV, respectively, which are much smaller than the measurement even though the experimental errors are considered. We notice that LHCb only used one Breit-Wigner distribution to depict the broad structure around 3188 MeV, which is still very rough. According to our result, two $\Omega_c^0$ states with $J^P=1/2^+$ and $J^P=3/2^+$ have the mass around 3.2 GeV, which mainly decay into $\Xi_c(2470)K$. It means that two $2S$ states should appear in the $\Xi_c(2470)K$ invariant mass spectrum simultaneously. Thus, we conjugate that such broad structure around 3188 MeV \cite{Aaij:2017nav} may contain at least two resonance structures corresponding to two $2S$ $\Omega_c^0$ states. If it is true, we may partly understand why the theoretical total decay widths of $2S$ $\Omega_c^0$ state with $J^P=1/2^+$ and $J^P=3/2^+$ are smaller than the experimental width of the $\Omega_c(3188)^0$. This conjugation should be further tested by experiment with more precise experimental data.

\begin{table}[htbp]
\caption{The partial and total decay widths in MeV, and branching fractions in \%, of the $2S$ $\Omega_c^0$ states. The partial widths of $D~\Xi$ are not listed since they are no more than 1 MeV for the $2S$ $\Omega_c^0$ states.} \label{table4}
\renewcommand\arraystretch{1.2}
\begin{tabular*}{85mm}{@{\extracolsep{\fill}}lcccc}
\toprule[1pt]\toprule[1pt]
Decay  &\multicolumn{2}{c}{$\Omega_c^0~[2S(1/2^+)]$}  & \multicolumn{2}{c}{$\Omega_c^0~[2S(3/2^+)]$}  \\
\cline{2-3}\cline{4-5}
modes  & $\Gamma_i$  & $\mathcal{B}_i$  & $\Gamma_i$  & $\mathcal{B}_i$ \\
\midrule[0.8pt]
 $\Xi_c(2470)K$        & 8.4     & 60.9\%   & 9.0       & 68.7\%      \\
 $\Xi^\prime_c(2570)K$ & 4.7     & 34.0\%   & 0.7       & 5.3\%      \\
 $\Xi^\ast_c(2645)K$   & 0.7     & 5.1\%    & 3.4       & 26.0\%    \\
 \midrule[0.8pt]
  Theory               & 13.8    & 100\%    & 13.1      & 100\%      \\
  Expt.~\cite{Aaij:2017nav} & \multicolumn{2}{l}{$60\pm15\pm11$} &   & \\
\bottomrule[1pt]\bottomrule[1pt]
\end{tabular*}
\end{table}

Here we present some typical ratios of partial decay widths for the $2S$ $\Omega_c^0$ states, $i.e$.,
\begin{equation}
\frac{\Gamma(\Omega_c^0(1/2^+)\rightarrow\Xi^\prime_c(2570)K)}{\Gamma(\Omega_c^0(1/2^+)\rightarrow\Xi_c(2470)K)}=0.56, \label{eq20}
\end{equation}
\begin{equation}
\frac{\Gamma(\Omega_c^0(1/2^+)\rightarrow\Xi^\ast_c(2645)K)}{\Gamma(\Omega^0_c(1/2^+)\rightarrow\Xi_c(2470)K)}=0.08, \label{eq21}
\end{equation}
\begin{equation}
\frac{\Gamma(\Omega_c^0(3/2^+)\rightarrow\Xi^\prime_c(2570)K)}{\Gamma(\Omega_c^0(3/2^+)\rightarrow\Xi_c(2470)K)}=0.08, \label{eq22}
\end{equation}
\begin{equation}
\frac{\Gamma(\Omega^0_c(3/2^+)\rightarrow\Xi^\ast_c(2645)K)}{\Gamma(\Omega_c^0(3/2^+)\rightarrow\Xi_c(2470)K)}=0.38, \label{eq23}
\end{equation}
which can be tested by future experiments. Finally, we should point out that $\Omega_c(3119)^0$ could not be an 2$S$ candidate in our scheme. As an $2S$ state with $J^P=1/2^+$ or $3/2^+$ , the total widths of $\Omega_c(3119)^0$ are predicted about 10.2 MeV and 8.3 MeV, respectively, which are much larger than the experimental result ($<2.6$ MeV, $95\% $ C.L.~\cite{Aaij:2017nav}).

\section{Summary and outlook}\label{sec5}

With the observation of five narrow $\Omega_c^0$ states at LHCb~\cite{Aaij:2017nav}, the study of higher orbital and radial excitations of charmed baryons is becoming a hot issue recently. These new $\Omega_c^0$ states also stimulated our interest in revealing their inner structures. By performing an analysis of their masses and strong decays, we find that five narrow $\Omega_c^0$ states, $i.e.$, $\Omega_c(3000)^0$, $\Omega_c(3050)^0$, $\Omega_c(3066)^0$, $\Omega_c(3090)^0$, and $\Omega_c(3119)^0$, could be grouped into the \emph{P}-wave charmed baryon family with $css$ configuration. Specifically, both $\Omega_c(3000)^0$ and $\Omega_c(3090)^0$ are the $J^P=1/2^-$ states with the mixtures of the $|0,1/2^-\rangle$ and $|1,1/2^-\rangle$, while the $\Omega_c(3050)^0$ and $\Omega_c(3119)^0$ are the $J^P=3/2^-$ states with the mixtures of the $|1,3/2^-\rangle$ and $|2,3/2^-\rangle$. Our results indicate that the mixing angles (see Eqs. (\ref{eq6}) and (\ref{eq7})) determined by the simple quark potential model are overestimated. In our scheme, the $\Omega_c(3066)^0$ is most like a $|2,5/2^-\rangle$ state. We also studied the possibility of the broad structure, $\Omega_c(3188)^0$~\cite{Aaij:2017nav}, as an $2S$ $\Omega_c^0$ state with $J^P=1/2^+$ or $J^P=3/2^+$. Our results suggest that such broad structure around 3188 MeV \cite{Aaij:2017nav} may contain at least two resonance structures corresponding to the $2S$ $\Omega_c^0$ states. Due to its very narrow decay width, in our scheme, $\Omega_c(3119)^0$ cannot be regarded an $2S$ candidate.

Since some important properties, such as the spin-parity quantum numbers, the EM transitions and the hadronic modes $\Omega_c^{(\ast)0}\pi$, have not been measured yet, other possible assignments may also exist for these five narrow $\Omega_c^0$ states. For examples, the measurements of these $\Omega_c^0$ states in the decay channels of $\Omega_c^{(\ast)}\pi$ may help us to test pentaquarks scenario~\cite{Kim:2017khv}. The EM transitions are also useful for providing important information about their internal structures~\cite{Wang:2017hej}. Even in our scheme, we can also explain the $\Omega_c(3050)^0$ and $\Omega_c(3090)^0$ states as the $1/2^-$ mixtures. Here the $\Omega_c(3050)^0$ state has the predominant $|1,1/2^-\rangle$ component while $\Omega_c(3090)^0$ has the predominant $|0,1/2^-\rangle$ component. Then the exotic assignment should be considered for the $\Omega_c(3000)^0$ state since a $1/2^-$ $\Omega_c^0$ state in the 1$P$ $\Omega_c^0$ family is suggested to have the lowest energy in most works (see discussion in Section \ref{sec2}). So an important task for the future experiments like LHCb and forthcoming BelleII is to carry out the measurement of the quantum numbers, the EM transitions, and other possible decay modes for these narrow $\Omega_c^0$ excitations.

Although the masses and widths of these 5 narrow $\Omega_c^0$ excitations could be explained under the $P$-wave assignment, at least two questions were not solved in the present work. The first one is why the value of $\gamma$ fixed in the $\Omega_c^0$ baryon family is so small. In our previous work~\cite{Chen:2016iyi}, a larger value of $\gamma$ in QPC model was fixed by the process of $\Sigma_c(2520)\rightarrow\Lambda_c~+~\pi$. Although the studying of meson decays indicates that the phenomenological parameter $\gamma$ in the QPC model may have a complicated structure~\cite{Segovia:2012cd,Bonnaz:1999zj}, this parameter have never been systematically investigated in the baryon sector. The second question is that why the mixing angles determined by the simple quark potential model are overestimated. To answer this question, more works are needed to investigate the spin-dependent interactions of these excited $\Omega_c^0$ states.

%\vspace{3ex}

%We have reason to believe that LHCb and the forthcoming BelleII will bring us more surprises.

\section*{Acknowledgement}

Bing Chen thanks Profs De-Min Li and Xian-Hui Zhong for the helpful suggestions. This paper is supported by the National Natural Science Foundation of China under Grant Nos. 11305003, 11222547, 11175073, 11447604, and 11647301. Xiang Liu is also supported by the National  Program for Support of Top-notch Young Professionals and the Fundamental Research Funds for the Central Universities.

\appendix
\section{Deduction of the flavor factors}\label{A}

We take the decay process, $\Omega_c^0\rightarrow\Xi_c^+~+~K^-$, as an example to show the evaluation of the flavor factors $\xi$ in Eq.~(\ref{eq18}). The flavor wave functions of initial and final states are given as below
\begin{equation}
\varphi_{\Omega_c^0}=ssc;~~~~~\varphi_{\Xi_c^+}=(us-su)c/\sqrt{2};~~~~~\varphi_{K^-}=s\bar{u}.\nonumber
\end{equation}
Then the flavor factors $\xi$ can be deduced by the following way (see Fig.~\ref{Fig2}),
\begin{equation}
\begin{split}
\xi&=\langle\varphi_{\Xi_c^+}\varphi_{K^-}|\varphi_{\Omega_c^0}\varphi_0\rangle\\
&=\left\langle\frac{(u_2s_5-s_2u_5)c_3}{\sqrt{2}}\times{s_1}\bar{u}_4\mid{s}_1s_2c_3\times\frac{u_5\bar{u}_4+d_5\bar{d}_4+s_5\bar{s}_4}{\sqrt{3}}\right\rangle\\
&=1/\sqrt{6}. \label{A1}
\end{split}
\end{equation}

Since the $s$ quarks in the initial excited $\Omega_c^0$ state have two possible ways to recombine in final states, we should time a statistical factor $\sqrt{2}$ for the value of flavor factor in Eq.~(\ref{A1}). The flavor factors $\xi$ of other decay processes related to this work can be calculated in the same way.

\section{The partial wave amplitudes for the decays of 1$P$ and 2$S$ $\Omega_c^0$ states}\label{B}

All partial wave amplitudes for the 1$P$ and 2$S$ $\Omega_c^0$ decays can be written as
\begin{equation}
\mathcal {M}_{LS}(q)=\mathcal{P}_{(\beta_i,m_j)}q^Le^{-\frac{4fh-g^2}{4f}q^2}.\label{B1}
\end{equation}
where the $\mathcal{P}_{(\beta_i,m_j)}$ are listed in Table~\ref{table5}. Here we have defined
\begin{equation}
\begin{split}
&f_s=\frac{6f\mu-(2fg\nu-g^2\mu)q^2}{16\sqrt{3}\pi^{5/4}f^{7/2}\lambda^{5/2}\beta_A^{5/2}\beta_{dA}^{3/2}\beta_{B}^{3/2}\beta_{dB}^{3/2}\beta_C^{3/2}};\\
&f_p=\frac{2f[12fg\lambda(1-\lambda\beta_A^2)+\mu(5g\mu-4f\nu)]+g(g\mu-2f\nu)^2p^2}{192\pi^{5/4}f^{9/2}\lambda^{7/2}\beta_A^{7/2}\beta_{dA}^{3/2}\beta_{B}^{3/2}\beta_{dB}^{3/2}\beta_C^{3/2}};\\
&f_d=\frac{g(g\mu-2f\nu)}{32\sqrt{5}\pi^{5/4}f^{7/2}\lambda^{5/2}\beta_A^{5/2}\beta_{dA}^{3/2}\beta_{B}^{3/2}\beta_{dB}^{3/2}\beta_C^{3/2}}, \label{B2}
\end{split}
\end{equation}
for the $s$, $p$, and $d-$wave decays. $\mu$, $\nu$, $\lambda$, $f$, $g$, and $h$ in Eqs.~(\ref{B1}) and (\ref{B2}) are given as
\begin{equation}
\begin{split}
&f=\frac{1}{2\beta_{dA}^2}+\frac{1}{2\beta_{dB}^2}+\frac{1}{2\beta_C^2}-\frac{\mu^2}{4\lambda};\\&
g=\frac{1}{\beta_{dA}^2}+\frac{\varepsilon_3}{\beta_{dB}^2}+\frac{\varepsilon_4}{\beta_C^2}-\frac{\mu\nu}{2\lambda};\\&
h=\frac{1}{2\beta_{dA}^2}+\frac{\varepsilon_2^2}{2\beta_B^2}+\frac{\varepsilon_3^2}{2\beta_{dB}^2}+\frac{\varepsilon_4^2}{2\beta_C^2}-\frac{\nu^2}{4\lambda};\\&
\lambda=\frac{1}{2\beta_A^2}+\frac{1}{2\beta_B^2}+\frac{\varepsilon_1^2}{2\beta_{dA}^2}+\frac{\varepsilon_3^2}{2\beta_{dB}^2};\\&
\mu=\frac{\varepsilon_1}{\beta_{dA}^2}+\frac{\varepsilon_3}{\beta_{dB}^2};~~\nu=\frac{\varepsilon_1}{\beta_{dA}^2}+\frac{\varepsilon_2}{\beta_B^2}+\frac{\varepsilon_3^2}{\beta_{dB}^2}.\nonumber
\end{split}
\end{equation}
$\varepsilon_1,~\cdots,~\varepsilon_4$ above are defined as
\begin{equation}
\begin{split}
&\varepsilon_1=\frac{m_1}{m_1+m_2};~~~~\varepsilon_2=\frac{m_3}{m_1+m_3+m_5};\\&
\varepsilon_3=\frac{m_5}{m_2+m_5};~~~~\varepsilon_4=\frac{m_4}{m_1+m_4}.\nonumber
\end{split}
\end{equation}
The SHO wave function scale parameters, $\beta_{d\alpha}$ ($\alpha=A$ or $B)$, reflect the sizes of diquark in the initial and finial baryons. $\beta_{\alpha}$ reflect distance between the light cluster and $c$ quark. $\beta_C$, reflects the sizes of the $K$ meson. For the 1$P$ and 2$S$ $\Omega_c^0$ states, the averaged values of the SHO wave function scale are obtained by the quark potential model, $i.e.$, $\beta_{1P}=0.236$ GeV and $\beta_{2S}=0.191$ GeV. The SHO wave function scale of light diquark, $\{ss\}$, is given as $\beta_{\{ss\}}=0.184$ GeV. The SHO wave function scales of other hadrons related in this work can be found in our previous work~\cite{Chen:2016iyi}. $m_j$ $(j=1,\cdots,5)$ denotes the quark masses in Fig.~\ref{Fig2}. In calculation of decays, the masses of $u/d$, $s$, and $c$ quarks are taken as 0.195 GeV, 0.380 GeV, and 1.680 GeV, respectively~\cite{Chen:2016iyi}.

\begin{table}[htbp]
\caption{The $\mathcal{P}_{(\beta_i,m_j)}$ for different decay modes of $1P$ and $2S$ $\Omega_c^0$ states.} \label{table5}
\renewcommand\arraystretch{1.4}
\begin{tabular*}{83mm}{@{\extracolsep{\fill}}lccc}
\toprule[1pt]\toprule[1pt]
$\mid j_l~J^P\rangle$            & $\Xi_c(2470)~K$                  & $\Xi^\prime_c(2570)~K$        & $\Xi^\ast_c(2645)~K$  \\
\midrule[0.8pt]
 $\mid 0,~1/2^-\rangle$  & $-\sqrt{1/2}f_S$         &   $\times$            &     \\
 $\mid 1,~1/2^-\rangle$  & $\times$                 &   $\sqrt{1/3}f_S$     &   \\
 $\mid 1,~3/2^-\rangle$  & $\times$                 & $-\sqrt{5/9}f_D$      &      \\
 $\mid 2,~3/2^-\rangle$  & $\sqrt{4/3}f_D$          & $f_D$                 &     \\
 $\mid 2,~5/2^-\rangle$  & $\sqrt{4/3}f_D$          & $-2/3f_D$             &      \\
 $\mid 1,~1/2^+\rangle$  & $\sqrt{1/2}f_P$          & $\sqrt{2/3}f_P$       & $\sqrt{1/3}f_P$  \\
 $\mid 1,~3/2^+\rangle$  & $\sqrt{1/2}f_P$          & $-\sqrt{1/6}f_P$      & $\sqrt{5/6}f_P$   \\
\bottomrule[1pt]\bottomrule[1pt]
\end{tabular*}
\end{table}

\vspace{3ex}

%\end{widetext}

\end{document}